# Rule based lexicographical permutation sequences

Asbjørn Brændeland

***Abstract***
In a permutation sequence built by means of sub permutations, the transitions between successive permutations are subject to a set of $n(n-1)/2$ rules that naturally group into $n-1$ matrices with a high degree of regularity. By means of these rules, the sequence can be produced in $O(3n!)$ time and $O(n^3)$ space.

To generate all permutations of a given set one can go for (among other things) a *minimum of changes* from one permutation to the next [1], [2], [4], or *lexicographic order*, at the cost of having to perform more changes [3]. Here, we go for the latter. A lexicographic order can be established by recursively producing successively larger sub-permutations, starting from the right, as in the following example on six elements. This makes the permuted elements move back and forth according to a set of *rules*, from one permutation to the next.

| | | | | | | | | | | | | | | |
|---|---|---|---|---|---|---|---|---|---|---|---|---|---|---|
| 1 | ( *a* | *b* | *c* | *d* | *e* | *f* ) | | 705 | ( *f* | *e* | *b* | *c* | *a* | *d* ) |
| 2 | ( *a* | *b* | *c* | *d* | *f* | *e* ) | | 706 | ( *f* | *e* | *b* | *c* | *d* | *a* ) |
| 3 | ( *a* | *b* | *c* | *e* | *d* | *f* ) | | 707 | ( *f* | *e* | *b* | *d* | *a* | *c* ) |
| 4 | ( *a* | *b* | *c* | *e* | *f* | *d* ) | | 708 | ( *f* | *e* | *b* | *d* | *c* | *a* ) |
| 5 | ( *a* | *b* | *c* | *f* | *d* | *e* ) | | 709 | ( *f* | *e* | *c* | *a* | *b* | *d* ) |
| 6 | ( *a* | *b* | *c* | *f* | *e* | *d* ) | | 710 | ( *f* | *e* | *c* | *a* | *d* | *b* ) |
| 7 | ( *a* | *b* | *d* | *c* | *e* | *f* ) | | 711 | ( *f* | *e* | *c* | *b* | *a* | *d* ) |
| 8 | ( *a* | *b* | *d* | *c* | *f* | *e* ) | | 712 | ( *f* | *e* | *c* | *b* | *d* | *a* ) |
| 9 | ( *a* | *b* | *d* | *e* | *c* | *f* ) | … | 713 | ( *f* | *e* | *c* | *d* | *a* | *b* ) |
| 10 | ( *a* | *b* | *d* | *e* | *f* | *c* ) | | 714 | ( *f* | *e* | *c* | *d* | *b* | *a* ) |
| 11 | ( *a* | *b* | *d* | *f* | *c* | *e* ) | | 715 | ( *f* | *e* | *d* | *a* | *b* | *c* ) |
| 12 | ( *a* | *b* | *d* | *f* | *e* | *c* ) | | 716 | ( *f* | *e* | *d* | *a* | *c* | *b* ) |
| 13 | ( *a* | *b* | *e* | *c* | *d* | *f* ) | | 717 | ( *f* | *e* | *d* | *b* | *a* | *c* ) |
| 14 | ( *a* | *b* | *e* | *c* | *f* | *d* ) | | 718 | ( *f* | *e* | *d* | *b* | *c* | *a* ) |
| 15 | ( *a* | *b* | *e* | *d* | *c* | *f* ) | | 719 | ( *f* | *e* | *d* | *c* | *a* | *b* ) |
| 16 | ( *a* | *b* | *e* | *d* | *f* | *c* ) | | 720 | ( *f* | *e* | *d* | *c* | *b* | *a* ) |

Here is one way of producing the sequence: Make a 6 × 6 matrix *M* with identical rows = [*a b c d e f*], take the downwards diagonal *D* and the 5 × 6 matrix $N = M \setminus D$, make the permutations of each row in *N* recursively and prepend $D[i]$ to each permutation of $N[i]$.

The principle is illustrated in Figure 1, with four instead of six elements.

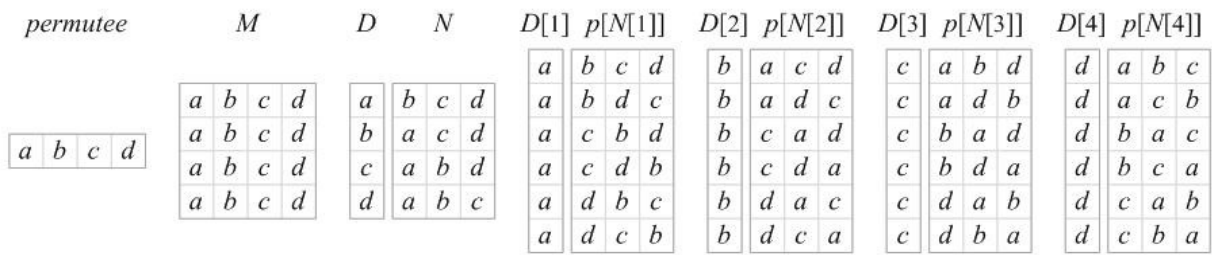

Figure 1.

The method, generalized to *n*-element sets, is easily implemented but not optimal. It requires more than $5(n-1)n!$ concatenations, and, since the entire set of sub-permutations must be in place before the final concatenations can be made, the space requirement is $n \cdot n!$. However, the number of permutation rules is only $\binom{n}{2}$ so if we can find these rules we can get an implementation that runs in $O(3n!)$ time and requires $O(n^3)$ space.

Let $P_i$ be the permutation sequence. The transition from for example $P_{12}$ to $P_{13}$ above is given by the rule [0, 0, 2, 2, –2, –2] which assigns the following moves to the elements $e_1, …, e_6$ in $P_{12}$: Leave $e_1$ and $e_2$ in place, move $e_3$ and $e_4$ two places to the right and move $e_5$ and $e_6$ two places to the left.

For a 6 elements permutation there are 15 rules. One set of rules brings a new element to the front. Since one of the 6 elements starts at the front there must be 5 such rules that apply to the permutations with indices 5!, 2·5!, 3·5!, 4·5! and 5·5!, and, except for these rules, for each successive segment of length 5! the rules must be the same. Within the first segment, a set of 4 rules brings a new element to position 2. These apply to indices 4!, 2·4!, 3·4! and 4·4! as well as to the corresponding indices in the subsequent segments of length 5!. Then there is a set of 3 rules that apply to indices 3!, 2·3!, 3·3!, and to the corresponding indices in the other length 4! segments, a set of 2 rules that apply to indices 2!, and 2·2!, in the length 3 segments and finally one rule that apply to index 1 and every other odd index.

We let the index of a rule $R$ be the index of the first permutation to which $R$ applies. It follows from the above reasoning that the permutations to which $R$ applies are evenly spaced, such that for $R_i$ the set of relevant indices is $i + j(i + 1)!$, for $j \geq 0$; and $(i + 1)!$ is then the ***step length*** of $R_i$.

All the rules in a set have the same step length. The step length $s$ for one set of rules $S_k$ equals the index of the first rule in the next set, $S_{k+1}$, and $s$ is also the distance between the rule indices of $S_{k+1}$. For $R_1$ the step length is 2, for $R_2$ and $R_4$ the step length is 6, for $R_6$, $R_{12}$ and $R_{18}$ the step length is 24, etc..

Now, we could reason about the contents of each rule, or we can simply observe them—and a striking pattern emerges: Each set of rules, leading zeros excluded, induces a matrix: a $1 \times 2$ matrix for the odd indices, a $2 \times 3$ matrix for the indices divisible by 2! but not by 3!, a $3 \times 4$ matrix for the indices divisible by 3! but not by 4!, etc. Let $RM_m$ be an $m \times m + 1$ matrix, let $C$ be the first column in $RM_m$, let $M$ be the $m \times m$ matrix $RM_m \setminus C$, and let $D$ be the upwards diagonal in $M$. Then $C[i] = i$, $D[i] = i - m - 1$, and otherwise $M[i, j] = m + 1 - 2j$, when $i$ and $j$ are the row and column indices, respectively.

| | | | | | | | | | |
|---|---|---|---|---|---|---|---|---|---|
| $R_1$ | [ 0 | 0 | 0 | 0 | 1 | –1 ] | 2 | { 1, 3, 5, 7, …, 717, 719} | 360 |
| $R_2$ | [ 0 | 0 | 0 | 1 | 1 | –2 ] | 6 | { 2, 8, 14, 20, …, 710, 716} | 120 |
| $R_4$ | [ 0 | 0 | 0 | 2 | –1 | –1 ] | 6 | { 4, 10, 16, 22, …, 712, 718} | 120 |
| $R_6$ | [ 0 | 0 | 1 | 2 | 0 | –3 ] | 24 | { 6, 30, 54, 78, …, 678, 704} | 30 |
| $R_{12}$ | [ 0 | 0 | 2 | 2 | –2 | –2 ] | 24 | {12, 36, 60, 84, …, 684, 710} | 30 |
| $R_{18}$ | [ 0 | 0 | 3 | –1 | 0 | –2 ] | 24 | {18, 42, 66, 90, …, 690, 714} | 30 |
| $R_{24}$ | [ 0 | 1 | 3 | 1 | –1 | –4 ] | 24 | {24, 144, 264, 384, 504, 624} | 6 |
| $R_{48}$ | [ 0 | 2 | 3 | 1 | –3 | –3 ] | 120 | {48, 168, 288, 408, 528, 648} | 6 |
| $R_{72}$ | [ 0 | 3 | 3 | –2 | –1 | –3 ] | 120 | {72, 192, 312, 432, 552, 672} | 6 |
| $R_{96}$ | [ 0 | 4 | –1 | 1 | –1 | –3 ] | 120 | {96, 216, 336, 456, 576, 696} | 6 |
| $R_{120}$ | [ 1 | 4 | 2 | 0 | –2 | –5 ] | 720 | {120} | 1 |
| $R_{240}$ | [ 2 | 4 | 2 | 0 | –4 | –4 ] | 720 | {240} | 1 |
| $R_{360}$ | [ 3 | 4 | 2 | –3 | –2 | –4 ] | 720 | {360} | 1 |
| $R_{480}$ | [ 4 | 4 | –2 | 0 | –2 | –4 ] | 720 | {480} | 1 |
| $R_{600}$ | [ 5 | –1 | 2 | 0 | –2 | –4 ] | 720 | {600} | 1 |

Figure 2. Rules, step lengths and the indices of the permutations to which the rules apply.

*Complexity*

The processing of one permutation involves searching for the pertinent rule and performing the moves. Since the search keys are constantly changed, as explained below, the average search time = $\binom{n}{2}/2$.

For each number of moves *m* there are *m* – 1 rules. Each 2-moves-rule applies to $n!/2!$ permutations, each 3-moves-rule applies to $n!/3!$ permutations, etc, and each *n*-moves-rule apply to $n!/n! = 1$ permutation. This gives the following number of moves:

$$1 \cdot 2 \cdot n!/2! + 2 \cdot 3 \cdot n!/3! + 3 \cdot 4 \cdot n!/4! + \ldots + (n-1) \cdot n \cdot n!/n!$$
$$= n!(1 + 1 + 1/2 + 1/2! + 1/3! + \ldots + 1/(n-2)!) \approx 3n!,$$

and the total time is about

$$n!(3 + \binom{n}{2}/2).$$

*Implementation*

For an actual generation of a permutation sequence we can represent each rule by an object containing the rule *R*, a permutation index *p* and a step length *s* and, to avoid an inordinate amount of rule tracking computation, we place the objects in some searchable structure. Together *p* and *s* give all the permutation indices to which *R* applies. Given permutation $P_p$, we look up the object *O* such that $O.p = p$ and produce $P_{p+1}$ by means of *O.R*, and <u>before we go to the next permutation, we set *O.p* to *O.p* + *O.s*</u>, so that *O* is searchable when the next permutation index to which *O.R* applies, comes around.

The implementation is written in *Racket*, a dialect of *Scheme*.

*Permutations by rules*

Since lists are immutable in Racket, and we need to update the rule index regularly, we use a vector rather than a list to represent the rule object.

```
(define (make-rule-object index step rule) (vector index step rule))
(define (get-index    rule-object) (vector-ref rule-object 0))
(define (get-step     rule-object) (vector-ref rule-object 1))
(define (get-rule     rule-object) (vector-ref rule-object 2))
(define (update-index! rule-object)
  (vector-set! rule-object
               0                                  ; The permutation index is the zero'th vector element.
               (+ (get-index rule-object)         ; Update the current index
                  (get-step rule-object))))       ; by adding the step length.
(define (move-elements n elements rule)
  (let ((perm-vector (make-vector n)))            ; Use a temporary vector for the new permutation.
    (for-each (lambda (elem move pos) (vector-set! perm-vector (+ pos move) elem))
              elements
              rule
              (enumerate 0 (- n 1)))              ; Current positions. †
    (vector->list perm-vector)))                  ; Return a list with the new permutation.
(define (make-perm-rule q i n)
  (define moves
          (cons i                                 ; The row indices fill the first column in the m × q matrix.
                (map (lambda (j)                  ; m × m matrix column index.
                       (if (= j (- q i))
                           (- i q)                ; [i,j] is in the upwards diagonal in the m × m matrix.
                           (- q (* 2 j))))
                     (enumerate 1 (- q 1)))))     ; m × m matrix column indices. †
  (front-pad-list 0 n moves))                     ; Fill inn leading zeros ‡
```

```
(define (make-perm-matrix m n)                          ; Permutation matrix index and permutee size.
  (define p (factorial m))                              ; Index of first rule in matrix.
  (define q (+ m 1))                                    ; Matrix width.
  (define s (* p q))                                    ; Step length.
  (map (lambda (p r) (make-rule-object p s r))
       (enumerate p (* p m) p)                          ; Permutation indices. †
       (map (lambda (i) (make-perm-rule q i n))         ; Rule.
            (enumerate 1 m))))                          ; Matrix row indices †
```

Here, we use a simple list for the permutation rules, which gives an average search time = $\binom{n}{2}/2$. Since the key values, the indices, are constantly changed, there are no obvious alternatives.

```
(define (set-up-perm-rules n)
  (flatten (map (lambda (m) (make-perm-matrix m n)) (enumerate 1 (- n 1)))))   ; †

(define (find-perm-rule perm-index rules)
  (let ((rule-object (first rules)))
    (if (= perm-index (get-index rule-object))          ; Found the right object, so
        (begin (update-index! rule-object)              ; prepare it for next search, and
               (get-rule rule-object))                  ; return the rule.
        (find-rule perm-index (rest rules)))))          ; Keep searching

(define (permute-by-rules permutee)
  (define n (length permutee))
  (define max-index (factorial n))
  (define rules (set-up-permutation-rules n))
  (define (iterate perm perm-index)
    (displayln perm)                                    ; Show the current permutation
    (if (>= perm-index max-index)
        'permutations-by-rules-completed
        (iterate (move-elements n perm (find-perm-rule perm-index rules))
                 (+ perm-index 1))))
  (iterate permutee 1))
```

---

; † `enumerate` takes two range arguments *a* and *b* and an optional step argument *s* and returns a list with the number sequence $a, a + s, a + 2s, …, a + ks$, where $a + ks \le b < a + (k + 1)s$. The default step value = 1.

; ‡ `front-pad-list` takes 3 arguments: a *pad*, a *full length*, and a *paddee*, and returns a list with the padde following a sufficient number of leading pads to fill the given length.